\documentclass[12pt,preprint]{aastex}

\newcommand{\nablaa}{\nabla_{\alpha}}
\newcommand{\nablab}{\nabla_{\beta}}
\newcommand{\pad}{\partial}
\newcommand{\md}{\mbox{d}}
\newcommand{\beq}{\begin{equation}}
\newcommand{\eeq}{\end{equation}}
\newcommand{\beqn}{\begin{eqnarray}}
\newcommand{\eeqn}{\end{eqnarray}}
\newcommand{\lppr}{\stackrel{<}{\scriptstyle \sim}}
\newcommand{\gppr}{\stackrel{>}{\scriptstyle \sim}}

\slugcomment{ApJ 617 (2004), 155}

\shorttitle{Shear acceleration in relativistic jets}
\shortauthors{}

\begin{document}


\title{Shear acceleration in relativistic astrophysical jets}


\author{Frank M. Rieger\altaffilmark{1} and Peter Duffy}
\affil{Department of Mathematical Physics, University College Dublin,
       Belfield, Dublin 4, Ireland}
\email{frank.rieger@ucd.ie; peter.duffy@ucd.ie}
\altaffiltext{1}{Marie Curie Postdoctoral Research Fellow (MCIF)}

\begin{abstract}
We consider the acceleration of energetic particles by a velocity shear
in the relativistic background flow containing scattering centers. Three 
possible acceleration sites for astrophysical jets are identified: (1) gradual 
velocity shear parallel to the jet axis, such as a velocity profile decreasing
linearly outward with radial coordinates, (2) gradual velocity shear perpendicular 
to the jet axis, such as intrinsic jet rotation, and (3) nongradual/discontinuous, 
longitudinal velocity shear at the jet-side boundary. We determine the 
characteristic acceleration timescales, specify the conditions for efficient 
acceleration, and discuss observational features with respect to each process. 
In particular, it is shown that in the case of (2) the higher energy emission 
is expected to be concentrated closer to the jet axis, while in the case of
(1) and (3) the higher energy particles are likely to be located near the edges 
of the jet, thus possibly leading to some form of limb-brightening.

\end{abstract}

\keywords{acceleration of particles --- galaxies: jets}

\section{Introduction}

Collimated relativistic outflows emerge from a variety of astrophysical sources
ranging from active galactic nuclei (AGNs), $\mu$-Quasars, and neutron stars in 
binary systems, to gamma-ray bursts \citep[]{fal97,zen97,mir99,fen04}.
It seems that in all of these sources the relativistic outflows, or at least 
those parts observed in the radio band, are essentially launched from an accretion 
disk \citep[]{liv99,mar02,fen04}. In several sources, monitoring of their 
radio structures clearly indicates outflowing components at high (apparently)
superluminal speeds \citep[]{mir99,jor01}, suggesting that the bulk plasma
in these jets moves at relativistic speed along the jet axis, with the jets being
observed at small viewing angles \citep[]{bla77}. The acceleration of particles 
and the origin of emission in such jets has been widely analyzed under the assumption 
of a one-dimensional flow profile. While clearly quite instructive, such an approach 
appears adequate only to a first approximation, since real jets are naturally expected 
to exhibit a significant velocity shear across the jet. In the present paper we 
therefore aim to explore the relevance of different shear velocity profiles for the 
acceleration of energetic particles in relativistic jets. The paper is structured 
as follows. In \S~2 basic results in the field of shear acceleration are reviewed. 
In \S\S~3 - 5 three possible sites of particle acceleration in astrophysical jets 
(e.g., velocity shear parallel and transversal to the jet direction) are distinguished 
on the basis of theoretical and observational arguments. We analyze their associated 
acceleration potential using simplified shear flow profiles and determine their 
characteristic timescales. Finally, possible consequences are addressed in \S~6.

\section{Physics and theory of shear acceleration}
Physically, particle acceleration in shear flows occurs as a consequence of energetic 
particles encountering different local flow velocities as they are elastically 
scattered off small-scale magnetic field inhomogeneities that are contained
in the collisionless, systematically moving background flow. In each 
scattering event the particle momentum is considered to be randomized in direction 
but its magnitude is assumed to be conserved in the (local) commoving flow frame 
in which the electric field vanishes. Since the rest frame of the scattering centers 
is regarded to be essentially that of the background flow (i.e., second-order Fermi 
effects due to a random component of motion of the scattering centres are neglected), 
particles will gain neither energy nor momentum merely by virtue of the scattering 
if there is no shear (or rotation) present and the flow is not diverging. However, 
in the presence of a shear in the background flow, the particle momentum relative to 
the flow changes for a particle travelling across the shear. As the particle momentum 
in the local flow frame is preserved in the subsequent scattering event, a net increase 
in particle momentum may occur \citep[]{jok90}.

The acceleration of energetic particles in collisionless plasmas with shear flows has 
attracted attention since the work of Berezhko \& Krymskii \citep[]
{berk81,ber81,ber82}. Starting with the nonrelativistic Boltzmann kinetic equation, 
they showed that the particle distribution function in the steady state might follow 
a power law, $ f \propto p\,'\,^{-(3+\alpha)}$, with $p\,'$ the magnitude of the 
commoving particle momentum, if the mean time interval between two scattering events 
increases with momentum according to a power law ($\tau \propto p\,'\,^{\alpha}$, 
$\alpha > 0$). The acceleration of cosmic rays at a gradual shear transition 
for non-relativistic flow velocities was studied independently by \citet{ear88}.
Assuming the distribution function to be almost isotropic in the commoving frame,
they rederived Parker's equation (i.e., the transport equation including the 
well-known effects of convection, diffusion and adiabatic energy changes) but also 
augmented it with new terms describing the viscous momentum transfer and the 
effects of inertial drifts. Effects of an average
magnetic field, not discussed by \citet{ear88}, were included in the paper by 
\citet{wil91}, showing that the magnetic field could lead to an anisotropic 
viscosity if the particle mean free path is not sufficiently small \citep[cf. also][]
{wil93}. \citet{jok90} presented a microscopic analysis of the particle transport in 
a moving, scattering fluid which undergoes a non-relativistic, step-function velocity 
change in the direction normal to the flow. They demonstrated that particles 
may gain energy at a rate proportional to the square of the velocity change.
Matching conditions in conjunction with Monte Carlo simulations for shear 
discontinuities were derived by \citet{jok89}. 

Particle acceleration using Monte Carlo methods for a sharp tangential velocity 
discontinuity has also been studied by \citet{ost90,ost98}. He showed that only 
relativistic flows are able to provide conditions for efficient acceleration, which 
could result in a very flat particle energy spectrum, depending only weakly on the 
scattering conditions. The possible relevance of such a scenario for the acceleration 
of particles at the transition layer between an AGN jet and its ambient 
medium was recently discussed by \citet{ost00} and \citet{sta02}. Astrophysical 
applications of shear acceleration were also presented recently by \citet{sub99}; 
following earlier suggestions by \citet{kat91}, they investigated the bulk acceleration 
of protons in the tenuous corona above an accretion disk, assuming the magnetic scattering 
centers (i.e., kinks in the magnetic field lines with the lines themselves being anchored 
in the disk) to be driven by the shear of the underlying Keplerian disk. They demonstrated 
that shear acceleration may transfer the energy required for powering the jet in blazar-type 
sources, accounting for bulk Lorentz factors up to $\sim 10$.

The work on gradual shear acceleration by \citet{ear88} was successfully extended to 
the relativistic regime by \citet{web89} \citep[cf. also,][]{web85}. Assuming the 
diffusion approximation to be valid, he derived the general relativistic diffusive 
particle transport equation for both rotating and shearing flows. Subsequently, Green's 
formula for the relativistic diffusive particle transport equation was developed by 
\citet{web94}. Applying their results to cosmic-ray transport in the galaxy, they 
found that the acceleration of cosmic rays beyond the knee by means of shear and 
centrifugal effects due to galactic rotation might be possible, but not to a sufficient 
extent. Using the relativistic transport theory developed by Webb~(1989), \citet{rie02} 
devised a simple model for the acceleration of energetic particles in rotating and 
shearing relativistic astrophysical jets. By considering possible velocity profiles 
for such jets, they analyzed the effects of shear and centrifugal acceleration and 
showed that (1) in the steady state, power-law particle spectra may be formed under 
a wide range of conditions, and (2) the inclusion of centrifugal effects may lead 
to a flattening of the spectra.\\ 
In this paper, we present for the first time an explicit derivation and analysis 
of the acceleration timescales in relativistic gradual shear flows and quantify the
conditions required for efficient particle acceleration. Observational and theoretical 
arguments are employed to distinguish between three possible acceleration sites and to 
show that shear acceleration of particles may be naturally expected in astrophysical 
jets.

\section{Longitudinal, gradual shear}
Generally, the interaction between a jet and the ambient medium is most likely to induce 
a velocity shear within the jet that is directed along the axis of propagation; of the 
type $\vec{v} = v_z(r)\,\vec{e}_z$. An energetic particle will experience a gradual shear 
profile if the velocity in the shear region changes continuously, and if its mean free 
path is much smaller than the transverse scale of the shear region. In the case of AGNs, 
for example, there is mounting observational evidence for jet stratification on the 
pc-scale, suggesting at least a two-component jet structure consisting of a fast inner 
spine (containing knots with transverse magnetic fields) and an extended, somewhat slower 
moving, shear boundary layer with a longitudinal magnetic field.  
Typical examples include the FR~II source 1055+018 \citep{att99}, the radio galaxies 3C353 
\citep{swa98} and M87 \citep{per99}, and the gamma-ray blazar Mrk~501 \citep{edw00}. 
The suggestions of a two-component jet structure are also supported by unification arguments 
showing that the radio and optical luminosities of BL Lac objects and FR~I radio galaxies 
are not consistent with predictions from simple unification schemes, such as within a one-zone 
model. It was shown by \citet{chi00}, however, that the results can be reconciled within the 
unifying scenario if one assumes an internal velocity structure in the jet, where a fast 
spine ($\Gamma_{\rm spine}\sim 15$) is surrounded by a slower, but still relativistically 
($\Gamma_{\rm layer}\sim 2$) moving, boundary layer. Relativistic hydrodynamical jet 
simulations may add further evidence revealing that the interaction between the jet and the 
ambient medium may indeed lead to a stratification of the jet in which a fast ($\Gamma_b 
\sim 7$) spine or beam is surrounded by a slower moving ($\Gamma_b \sim 1.7$), high-energy 
shear layer \citep[cf.][]{alo00,gom02}. The shear layer in these simulations has been found 
to broaden with distance along the jet, reaching a transverse size comparable to the beam radius. 
There is also increasing evidence for jet stratification on larger scales. In order to obtain 
a good qualitative description of the total intensity and polarization systematics of the jets 
in a sample of FR I sources, \citet{lai96} \citep[cf. also][]{har97} has suggested, 
that FR~I jets consist of a fast spine with transverse magnetic fields, which decelerates from 
relativistic $\beta_s$ to nonrelativistic speeds on kpc scales, and a slower moving shear 
layer with a longitudinal or two-dimensional magnetic field, where the velocity decreases linearly
with radial distance from the jet axis, which is probably caused by entrainment of external 
material. More specifically, it appears that the anticorrelation between brightness ratio (of 
jet to counter-jet) and width ratio from Gaussian fitting in the kpc-jets of FR I radio galaxies 
is suggestive of a relativistic flow profile that decreases radially outward from the jet axis, 
with the ratio $\beta_{\rm z,max}/\beta_{\rm z,min}$ of velocities
between the center (the fast spine) and the slower moving shear layer (the edge of the jet) 
required to be in the range $\sim (0.1-0.2)$ in order to produce the observed anticorrelation 
\citep{lai99}.

In order to analyze the conditions for efficient particle acceleration in such shear flows,
let us consider a simple background flow that moves at relativistic speed along the jet 
axis with magnitude depending on the radial coordinate $r$, so that in four-vector notation 
the bulk flow profile is of the form
\beq
           u^{\alpha}= \gamma_b\,\left(1,0,0,v_z(r)/c\right)
\eeq where $\gamma_b(r)=1/\sqrt{1-v_z(r)^2/c^2}$ denotes the bulk Lorentz factor. 
In the commoving frame the shear acceleration coefficient may be written as \citep[e.g.,][]
{web89}
\beq
   \left<\frac{\Delta p\,'}{\Delta t}\right> = \frac{1}{p\,'^{\,2}}\frac{\pad }{\pad p\,'}
                \left(p\,'^{\,4}\,\tau\,\Gamma\right)
\eeq where $p\,'$ is the commoving particle momentum and $\tau \simeq \lambda/c$ is the mean 
time between two scattering events. For the above flow profile and by assuming isotropic 
diffusion, the relativistic shear coefficient $\Gamma$ is given by (cf. eq.~[\ref{shear-coeff}])
\beq\label{shear-coeff}
        \Gamma = \frac{1}{15}\,\gamma_b(r)^2 \left(\frac{\pad v_z(r)}{\pad r}\right)^2\,
                 \left[1 + \gamma_b(r)^2\,\frac{v_z(r)^2}{c^2}\right]\,,
\eeq and hence the characteristic acceleration timescale in the commoving frame becomes
\beq
      t_{\rm sh,l}(r)  \simeq  \frac{p\,'}{<\dot{p}\,'>}  =  \frac{15}{(4+\alpha)}\,
            \frac{c}{\lambda}\,\frac{1}{\gamma_b(r)^2\,\left(\frac{\pad v_z}{\pad r}\right)^2
                      \left[1+\gamma_b(r)^2\,\frac{v_z^2}{c^2}\right]}
\eeq where $v_z \equiv v_z(r)$ and a power law energy dependence for the scattering timescale
has been assumed, $\tau \propto p\,'\,^{\alpha}$. 

Possible realizations of longitudinal shear flows include a velocity shear profile decreasing 
linearly outward \citep[]{lai99} and a top-hat - type velocity profile as suggested from
hydrodynamical simulations. The former can be modeled as a velocity profile with constant inner 
spine, i.e. relativistic $v_{\rm z,max}\simeq c$ inside $r_1$ and nonrelativistic $v_{z,\rm min}$ 
at $r_2 > r_1$, where the size of $r_2$ may be typically taken to be of the order of the jet 
radius so that
\beq
          v_z(r) = v_{\rm z,max} - \frac{\Delta v_z}{\Delta r}\,(r-r_1)\,H(r-r_1)\,,
\eeq where $H(r-r_1)$ denotes the Heaviside step function, $\Delta v_z = (v_{\rm z,max} - 
v_{\rm z,min}) \sim c$ and $\Delta r = (r_2 - r_1)$. The associated acceleration timescale is 
then of the order of
\beq     
    t_{\rm sh,l}(r) \simeq \frac{3}{c\,\lambda}\,\frac{(\Delta r)^2}{\gamma_b(r)^2\,
                     \left[1+\gamma_b(r)^2\,\frac{v_z(r)^2}{c^2}\right]}
\eeq
It seems clear that for efficient acceleration of electrons a shear layer of sufficient small 
radial extension is required. For typical relativistic, pc-scale radio properties in AGN jets
[$B \sim 0.01$ G, $\gamma_b(r_1) \sim  10$, $\gamma \sim 300$ for radio emitting electrons 
and hence $\lambda \lppr r_g \sim 5 \cdot 10^7\,(\gamma/300)$ cm], we
may then have an electron acceleration timescale near $r_1$ of order
\beq\label{estimate1}
    t_{\rm sh,l} \sim \frac{3\;(\Delta r)^2}{\gamma_b^4\,c\,r_g} 
                 \simeq 500\,\left(\frac{300}{\gamma}\right)\,
                  \left(\frac{\Delta r}{0.003\,{\rm pc}}\right)^2\;\; {\rm yrs}\,,
\eeq while the corresponding isotropic electron synchrotron cooling timescale is of the order
$t_{\rm cool} \sim  800 \,(300/\gamma)\,(0.01\,{\rm G}/B)^2$ yrs. 
Hence, for efficient electron acceleration one would need a $\Delta r$ of the order of 
$0.001$ pc or less. However, the conditions are much more favorable for proton acceleration, 
where (for producing emission at the same frequency, and all else being equal) $\Delta r$ could 
be a factor $m_p/m_e$ larger than the one required for efficient electron acceleration, so that 
for protons $\Delta r$ may be of the order of $1$ pc. This suggests that under general 
conditions shear acceleration of protons resulting in proton synchrotron emission is more 
likely to occur, even though usually the emission by protons may be neglected because of its
relatively small emitted power. It could be shown that a similar conclusion 
holds for a top-hat velocity profile with a fast core region, an adjacent transition layer, and a 
slower outer collar extension, except that theoretically the transition layer no longer needs to 
be located close to the jet-side boundary. Generally, it is interesting to note that the shear 
acceleration timescale depends inversely on the particle mean free path, $t_{\rm sh,l} 
\propto 1/\lambda$, or $t_{\rm sh,l} \propto 1/\gamma$ if we assume a gyroradius momentum 
dependence. The synchrotron loss timescale shows the same proportionality, i.e., $t_{\rm cool} 
\propto 1/\gamma$; hence once shear acceleration works efficiently, synchrotron losses are no 
longer able to stop the acceleration, and particles are continuously accelerated to higher 
energies unless a process like escape or cross-field diffusion becomes dominant. 

\section{Longitudinal, non-gradual shear}
We now consider a particle that becomes so energetic that its mean free path becomes larger than
the transversal width of the velocity transition layer. The particle can then be regarded 
as passing unaffected through this layer and hence essentially experience a discontinuous drop 
in velocity. If the particle elastically scatters off magnetic inhomogeneities frozen into the 
background flows and repeatedly crosses the layer, efficient particle acceleration to high 
energy may occur. \citet{ost90}, for example, has argued that the interface between the jet 
interior in a powerful AGN and the ambient medium may represent such a transition layer across 
which a relativistic velocity drop naturally occurs. We therefore consider 
a velocity transition layer between two fluids A and B, where fluid A moves along the jet axis 
with velocity $u_1$ and fluid B with velocity $u_2$. Relativistic particles with energy $E_0$ 
measured in the commoving frame A, which pass unaffected through the transition layer of width
$\Delta r$, will have a total energy in the commoving frame B given by the Lorentz transformation
\beq\label{energy}
           E'=\Gamma_{\Delta}\,E_0\,(1 + \beta\,\cos\phi) \simeq \Gamma_{\Delta}\,E_0 \,,
\eeq where $\cos\phi = \vec{p}\cdot\Delta\vec{u}/[\,|\vec{p}|\cdot|\Delta\vec{u}|\,]$ with
$\vec{p}$ the particle momentum, $\Delta u = (u_1-u_2)/(1-u_1\,u_2/c^2)$ the relative velocity 
difference, $\beta=\Delta u/c$ and $\Gamma_{\Delta}=1/\sqrt{1-(\Delta u/c)^2}$ the associated
Lorentz factor. The second relation on the rhs of equation~(\ref{energy}) holds if the particle 
distribution is assumed to be almost isotropic near the velocity discontinuity. Hence, when a 
particle crosses the transition layer from A to B and elastically scatters off magnetic 
inhomogeneities frozen in the background flows, its energy has this value in the local rest 
frame B. If the particle recrosses the transition layer (from B to A), its total energy 
measured in the commoving frame A is then
\beq
           E = \Gamma_{\Delta}\,E'\,(1-\beta\,\cos\phi') \simeq \Gamma_{\Delta}\,E'
             = \Gamma_{\Delta}^2\,E_0\,,
\eeq where the second relation on the rhs again holds for an almost isotropic particle 
distribution. After two successive crossings, particles may thus increase their energy by a 
factor of $\Gamma_{\Delta}^2$. Hence, if the particle distribution is almost isotropic near 
the discontinuity, the mean energy gain for a single crossing of the discontinuity
is given by (cf. eq.~[\ref{energy}])
\beq
     \left< \frac{\Delta E}{E}\right> \simeq (\Gamma_{\Delta} - 1)\,.
\eeq It seems clear that according to this idealized picture, the increase in particle energy 
may be substantial provided the velocity difference is relativistic. For nonrelativistic 
velocities, however, only a mean energy gain of second order in $\Delta u$ is 
obtained. In the presence of relativistic flow speeds the full picture requires that the 
nonnegligible anisotropy of the particle distribution is taken into account. The mean energy
gain may then be expressed as
\beq
        \left< \frac{\Delta E}{E}\right> =\eta_e\, (\Gamma_{\Delta} - 1)\,,
\eeq with $\eta_e < 1$. Using Monte Carlo particle simulations within the strong scattering 
limit ($\Delta B/B \sim 1$, $\kappa_{\perp} \sim \kappa_{||}$), \citet{ost90} has shown 
that $\eta_e$ may still be a substantial fraction of unity. Technically, one can define a 
characteristic acceleration timescale given by $t_{\rm acc} = <\Delta t>/<\Delta E/E>$, where 
$<\Delta t>\,\propto\, \tau$ is the mean time for boundary crossing and $\tau = \lambda / c$, 
with $\lambda$ the particle mean free path. In the ambient medium rest frame one may thus finally 
arrive at \citep[cf.][]{ost98,ost00}
\beq
   t_{\rm acc} = \alpha \;\, \frac{\lambda}{c}\;\; \gppr \;(1-10) \; \frac{r_g}{c} \quad
                        {\rm if} \;\; r_g >  \Delta r \,\,,
\eeq where particle simulations suggest that $\alpha$ may be as small as $\sim (1-10)$, provided
that particles are allowed to escape from the acceleration process once they have crossed a boundary 
at $r_{\rm max}$. Note that $\alpha$ is, of course, dependent on the radial escape boundary distance
$r_{\rm max}$, so that for $\Delta u/c=0.99$ and $r_{\rm max}/r_g =3,9,18$, for example, one may 
have $\alpha \simeq 4,9,18$ \citep[cf.][Fig.~3a]{ost90}. In the presence of a longitudinal mean 
magnetic field, the particle mean free path may be of the order of the gyroradius $r_g$ 
\citep[cf.][]{ost98,sta02}, but should be still larger than the width $\Delta r$ of the transition 
layer, i.e., the acceleration mechanism may work fairly quick, provided we have very energetic 
particles with $\lambda \sim r_g > \Delta r$ and $r_g < r_j$, where $r_j$ denotes the jet radius. 
Observations of pc-scale jets with evidence for a shear layer morphology (e.g., a boundary 
layer with parallel magnetic fields) suggest that $\Delta r < 0.5\,r_j$. Taking $\Delta r \sim 
0.1\,r_j \sim 0.1$ pc and $B \sim 0.01$ G for a semi-quantitative estimate, we would require very 
high seed electron Lorentz factors of about $\gamma_e \sim 10^{12}$ and proton Lorentz factors 
of about $\gamma_p \sim 5 \cdot 10^8$. The acceleration timescale should then be of the order of 
$t_{\rm acc} \gppr 0.1 \,{\rm pc}/c \sim 0.3$ yrs. Yet, due to the rapid radiation losses for 
electrons at the required high energies, the considered mechanism is unlikely to work efficiently
for electrons since one typically expects $r_g(e) \ll \Delta r$. Again the mechanism is much more 
favorable for protons; it may be quite possible to accelerate protons efficiently 
until their gyroradius becomes larger than the width of the jet. Note that for this case we have 
$t_{\rm acc}\propto \lambda$, while for the gradual shear cases above $t_{\rm sh,l} \propto 1/
\lambda$.

\section{Transversal, gradual shear}
In addition to the possible occurrence of a longitudinal velocity shear in astrophysical jets, 
real jets are also likely to exhibit a velocity shear transversal to the jet axis. In particular, 
several independent arguments suggest that jets may generally be characterized by an additional 
rotational velocity component. The analysis of the jet energetics in extragalactic radio 
sources for example has revealed a remarkably universal correlation between a disk luminosity
indicator and the bulk kinetic power in the jet \citep[]{raw91,cel93} and established a close 
link between jet and disk \citep[cf. also][]{liv99,mar02}, a fact which has been successfully 
exploited within the concept of a so-called jet-disk symbiosis \citep[e.g.,][]{fal95a,fal95b}.
The results suggest that a significant amount of accretion energy, and hence rotational energy 
of the disk (cf. virial theorem), is channeled into the jet allowing for an efficient removal 
of angular momentum from the disk. Furthermore, direct evidence for intrinsic jet rotation has 
been recently established for stellar outflows \citep[e.g.,][]{bac00,bac02,dav00,wis01,tes02,cof04}.  
The optical jet in the T Tauri star DG Tau (central mass $M_{\star} \simeq 0.67\,M_{\odot}$),
for example, is associated with a Keplerian rotating circumstellar disk and appears to have an
onionlike structure, with the faster and more collimated gas continuously bracketed by wider 
and slower material. The flow becomes gradually denser toward the central axis. High angular 
resolution observations revealed a systematic offset in the radial velocities of optical emission 
lines found in pairs of slits on alternate side of the jet axis, with average Doppler 
shifts being in the range of $5-20$ km/s. This suggests a rotating jet flow with intrinsic,
position-dependent, toroidal velocities in the range $(6-15)$ km/s \citep[for possible 
underestimation-effects, see however][]{pes04} in the region probed by the observations, i.e. 
within several tens of AU from the jet axis \citep[cf.][]{bac00,bac02,tes02}. 
In the case of AGNs intrinsic jet rotation is also suggested by several indirect 
observations such as the helical motion of knots, the apparently 
oscillating ridge line in pc-scale jets, the detection of double helical patterns and the 
observations of periodic variabilities \citep[e.g.,][]{cam92,sch93,ste95,wag95,kri00,lob01}. 
In the lighthouse model proposed by \citet[]{cam92}, the origin of periodicity has 
been successfully related to differential Doppler boosting of emission from an off-axisymmetric 
density perturbation in the parsec-scale jet, which is dragged along with the underlying 
rotating bulk of the plasma.
Finally, from a more theoretical point of view, intrinsic jet rotation is generally expected 
in magnetohydrodynamic (MHD) models for the formation and collimation of relativistic
astrophysical jets \citep[e.g.,][]{beg94,cam96,fen97,fer98,sau02}. In such models, intrinsic 
rotation with speeds up to a considerable fraction of the velocity of light is a natural 
consequence of the assumption that the flow is centrifugally accelerated from the accretion 
disk. The set of available rotation profiles needs not necessarily to be disklike 
\citep[i.e., Keplerian, cf.][]{bla82}, but may also include, for example, rigid and flat 
azimuthal velocity profiles \citep[e.g.,][]{vla98,han00,ler00}.

In order to analyze the capability of particle acceleration by the shear in rotating flows, 
let us consider a simple model, in which the relativistic outflow velocity $v_z$ along the jet 
axis is kept constant and the velocity component perpendicular to the jet axis is purely 
azimuthal, described by the angular frequency profile $\Omega(r)$. In the local 
commoving frame, the characteristic shear acceleration timescale for energetic particles with 
$v'\simeq c$ may then be estimated using \citep[e.g.][]{web89,rie02}
\beq\label{azimu_shear}
    t_{\rm shear} \simeq \frac{p\,'}{<\dot{p}\,'>_s}
                   =\frac{c}{(4+\alpha)\,\lambda\,\Gamma}
\eeq where the shear coefficient $\Gamma$ is given by \citep[cf.][]{rie02}
\beq
    \Gamma= \frac{1}{15} \,\gamma_b(r)^4\,r^2\,\left[\frac{\md \Omega(r)}{\md r}\right]^2
              \,(1-v_z^2/c^2)\,.
\eeq Here $\gamma_b(r) = (1-v_z^2/c^2-\Omega(r)^2\,r^2/c^2)^{-1/2}$ denotes the position-dependent
Lorentz factor of the background flow, $\lambda \equiv \lambda(\gamma) =\lambda_0\,\gamma^{\alpha} 
< r_g$, with $\alpha > 0$, is the particle mean free path, and $\gamma$ is the Lorentz factor of 
the particle. Since the jet flow is likely to be launched from an accretion disk, one expects the 
jet flow to be radially confined to a region between $r_{\rm in}$ and $r_{\rm out} \simeq r_j$. 
Keplerian or flat rotation belong to the most relevant realizations of azimuthal shear profiles, 
and we will thus center our discussion here on these two types:\\
(i) For a Keplerian rotation profile of the form $\Omega(r)=\tilde{\Omega}_k \,(r_{\rm in}/
r)^{3/2}$, where $\tilde{\Omega}_k$ is a constant and $r_{\rm in} \leq r \leq r_j$, 
Eq.~(\ref{azimu_shear}) results in
\beq
    t_{\rm shear}^k(r) \simeq \frac{20}{3\,(4+\alpha)}\;\frac{c}{\lambda}\,\frac{1}{\gamma_b^4}\,
                   \frac{1}{(1-v_z^2/c^2)}\,\frac{1}{\tilde{\Omega}_k^2}\,
                   \left(\frac{r}{r_{\rm in}}\right)^3\,.
\eeq For a quantitative estimate we assume that because of a net radial mass flow inward, 
the rotational profile remains similar when the jet boundaries open slightly with 
distance, so that the rotational velocity at the inner boundary of the jet is of the order of the
rotation velocity near the innermost stable orbit $r_{\rm is}$ of the underlying accretion disk.
We then have $\tilde{\Omega}_k \simeq v_k(r_{\rm in})/r_{\rm is}$ with $r_{\rm is}=k_0\,r_s$,
where $r_s$ denotes the Schwarzschild radius, and $k_0 = 0.5$ in the Kerr BH case, so that the
acceleration timescale becomes
\beq
    t_{\rm shear}^k(r) \simeq \frac{20}{3\,(4+\alpha)}\;\frac{c}{\lambda}\,\frac{1}{\gamma_b^4}\,
                        \frac{1}{(1-v_z^2/c^2)}\;\frac{1}{v_k(r_{\rm in})^2}\;k_0^2\;r_s^2\;
                         \left(\frac{r}{r_{\rm in}}\right)^3\,.
\eeq Using physically reasonable parameters, i.e., $r_{\rm in} \simeq 0.01$ pc, $v_k(r_{\rm in}) = 
\xi_k\,c$ with $\xi_k \leq (1-v_z^2/c^2)^{1/2}$ $\simeq 0.3$ for $v_z/c \simeq 0.95$, $k_0=0.5$ 
(Kerr BH), and $\lambda(\gamma) = \eta\,r_g$, $r_g \sim 5 \times 10^7\,(\gamma/300)$ cm, $\eta 
< 1$, $\alpha = 1$, $r_s \simeq 3 \cdot 10^{13}$ cm (i.e. $M_{\rm BH} \sim 10^8\,M_{\odot}$)
and $\gamma_b \sim 4$, one can obtain an order of magnitude estimate for the acceleration of radio 
emitting electrons in pc-scale jets given by 
\beq
    t_{\rm shear}^k(r) \sim 30 \,\left(\frac{300}{\gamma}\right) 
                                 \left(\frac{r}{0.01\,{\rm pc}}\right)^3\;\;{\rm  yrs}\,. 
\eeq The corresponding (isotropic) electron synchrotron cooling timescale, on the other hand, is 
of the order $t_{\rm cool} \sim 800 \,(300/\gamma)\,(0.01\,{\rm G}/B)^2$ yrs. It seems clear 
that (a.) for electrons the acceleration process is rather limited and confined to a small inner
region of the jet, while it is again much more favorable for protons, where (for producing emission 
at the same frequency, and all else being equal) $t_{\rm shear, p}^k(r) \sim 0.7\, (r/0.01\,
{\rm pc})^3$ yrs, while $t_{\rm cool, p} \sim 3\cdot 10^7$ yrs, and (b.) substantial rotation 
is required for efficient acceleration.\\
(ii) For a flat rotation profile of the form $\Omega(r)=\tilde{\Omega}_f \,(r_{\rm in}/r)$ on the
other hand, where $\tilde{\Omega}_f$ is a constant and $r_{\rm in}\leq r \leq r_j$, 
equation.~(\ref{azimu_shear}) becomes
\beq
    t_{\rm shear}^f(r) \simeq \frac{15}{(4+\alpha)}\, \frac{c}{\lambda}\,
                      \,\frac{1}{\gamma_b^4}\,\frac{1}{(1-v_z^2/c^2)}\,\frac{1}{\tilde{\Omega}_f^2}
                      \,\left(\frac{r}{r_{\rm in}}\right)^2\,.
\eeq For a quantitative estimate, we again suppose maintenance of the rotational profile, 
so that $\tilde{\Omega}_f \simeq v_{\phi,f}/r_{\rm is}$ where $v_{\phi,f}=\xi_f\,c$, with $\xi_f 
\leq \sqrt{1-v_z^2/c^2} \simeq 0.3$. Using the same parameters as above (i.e., $r_{\rm is}=k_0\,r_s$ 
with $k_0=0.5$, $\gamma_f \sim 4$ etc.), we may obtain an order of magnitude estimate for the 
acceleration of electrons by shear given by 
\beq 
    t_{\rm shear}^f(r) \sim 3\, \left(\frac{300}{\gamma}\right)
                                 \left(\frac{r}{0.01\,{\rm pc}}\right)^2\;\;{\rm yrs}\,.
\eeq The acceleration potential is then somewhat improved compared to the Keplerian case, but 
it is still rather limited for electrons, while it is again much more favorable for protons.\\  
Our present analysis is aimed toward a study of the shear acceleration characteristics, and hence 
we have neglected centrifugal acceleration effects in rotating flows. We note however, that such 
effects may become particularly relevant for flat rotation profiles, in which the centrifugal 
acceleration timescale could be of the same order of magnitude as the shear timescale 
\citep[cf.][]{rie02}.
Generally, the shear acceleration timescale for rotating flows increases with the radial coordinate 
(i.e., with the cube for the keplerian case and the square for the flat case), and hence the higher 
energy emission is naturally expected to be concentrated closer toward the jet axis. As the 
synchrotron loss timescale scales with $1/\gamma$, synchrotron losses are again no longer able 
to stop the acceleration once the shear mechanism has started to work efficiently (cf. \S~3).

\section{Discussion and conclusion}
In the present paper we have identified three possible velocity shear sites in powerful 
astrophysical jets on the basis of theoretical and observational arguments. It seems 
intuitively clear that the scattering of energetic particles across such shear profiles 
should naturally lead to a net increase in particle momentum, i.e., to the acceleration of 
particles. Using an idealized approach, we have specified order-of-magnitude estimates for 
the characteristic acceleration timescales using reasonable parameters for AGN jets. A more 
detailed determination of the acceleration efficiency, which seems clearly desirable, however 
would require a detailed, but currently inaccessible knowledge of the real intrinsic jet 
properties and a more rigorous treatment of the physical mechanism responsible for the 
scattering. The last point is particularly illustrated by the dependence of the acceleration 
timescale on the particle mean free path (inversely proportional for the gradual shear 
cases and proportional for the nongradual shear) and, in the presence of a longitudinal mean 
magnetic field, by the need of a sufficiently strong diffusion perpendicular to the field. Our 
present approach for the gradual shear, for example, is based on an idealized large angle (hard 
sphere) scattering model, where $\kappa_{\perp} \simeq \kappa_{||}$ in the limit $\omega_g'\,
\tau \ll 1$. While we thus caution that more detailed physical models need to be developed 
before further conclusions can be drawn, we nevertheless suggest that the generic nature of 
our analysis might give valuable insights into the acceleration of particles by the shear in 
astrophysical jets.\\
In general, the particle mean free path for electrons is much smaller than the particle mean 
free path for protons emitting at the same frequency. Shear acceleration is thus much more 
favorable for protons than for electrons. For the considered realizations, balancing the 
acceleration timescales with synchrotron losses indeed suggests that shear acceleration is 
rather marginally significant for electrons, while it could be reasonably efficient for 
protons.\\ 
Phenomenologically, one expects that particle acceleration by a (nongradual) shear at the 
jet-side boundary and electron acceleration by a (gradual) longitudinal velocity shear within
the jet, which decreases linearly with the radial coordinate, are associated with the higher 
energy particles being concentrated closer toward the edges of the jet, and thus possibly 
related to some form of limb brightening given suitable flow velocities, jet orientations, and 
magnetic field strengths. The picture is, however, generally more complex if gradual shear 
acceleration of protons or top-hat velocity profiles are considered. For example, if a 
top-hat-type velocity profile with a radially extended collar is indeed physically realized, 
as perhaps suggested by two-flow models \citep[]{sol89,rol94}, the situation may be 
reversed, such that higher energy particles are located closer toward the axis. On the other 
hand, in the case of an intrinsically rotating flow, the shear acceleration timescale usually 
increases strongly with radial coordinate, so that higher energy particles are generally 
expected to be concentrated closer towards the jet axis. It is, however, again possible, that 
the real, observed situation is more ambiguous. In the case of an internally rotating jet flow, 
for example, a transition layer from the rotating jet interior to the nonrotating environment is 
still necessary. For a comparatively weak magnetic field and a sufficiently strong rotationally 
sheared layer this may lead to the onset of a strong magnetorotational instability \citep[cf.]
[]{bal91}, resulting in the growth of small-scale, tangential magnetic field discontinuities 
until the magnetic energy is eventually dissipated through reconnection events in which efficient 
particle acceleration may occur. In that case we may also have a contribution of highly energetic 
particles from the edges of the jet.\\
So far, we have not considered the relevance of second-order Fermi effects that may contribute 
to the acceleration of particles \citep[for a complementary approach, see][]{man99,sta02}. Yet,
if the scattering centers are not assumed to be completely frozen in
the background flow, second-order Fermi effects should become dominant at lower energies. 
Quantifying these effects by determining the intrinsic Alfven velocity, for example, is, however, 
notoriously difficult because of our ignorance of the real physical parameters in astrophysical 
jets. Nevertheless, it seems qualitatively evident that, for a gyroradius momentum dependence of the 
particle mean free path, the timescale for second-order Fermi acceleration scales as $t_{\rm 2nd\,F} 
\propto \lambda  \propto  \gamma$. The acceleration timescales in the gradual shear cases on the
other hand, scales as $t_{\rm sh} \propto 1/ \lambda \propto 1/\gamma$. Hence, even if the observed 
lower energy emission is mainly due to particles accelerated via a second-order Fermi process, there 
exists an energy scale above which gradual shear acceleration becomes dominant. In that case, 
second-order Fermi acceleration may be considered as providing the seed particles for efficient 
gradual shear acceleration within the jet. Accordingly, one may perhaps develop a picture in which 
supra-thermal particles, injected and centrifugally accelerated at the base of a jet \citep[e.g.,]
[]{rie00}, are -- in the absence of strong shocks -- initially further accelerated along the jet 
via second-order Fermi and then subsequently via gradual shear effects, until eventually 
(nongradual) shear acceleration at the jet-side boundary becomes possible, potentially allowing 
for the formation of a cosmic-ray cocoon \citep[cf.][]{ost00}.

\acknowledgments
We are grateful to M.~Ostrowski and K.~Mannheim for discussions and comments.
Useful comments by an anonymous referee, which helped to improve the presentation, 
are also gratefully acknowledged.
FMR acknowledges support through a Marie-Curie Individual Fellowship 
(MCIF-2002-00842).

\appendix
\section{Longitudinal, gradual shear flow}
In the commoving frame, the acceleration coefficient due to a velocity shear can be 
cast into the form \citep[e.g.,][eq.~3.27]{web89}
\beq
   \left<\frac{\Delta p\,'}{\Delta t}\right> = \frac{1}{p\,'^{\,2}}\frac{\pad }{\pad p\,'}
                \left(p\,'^{\,4}\,\tau\,\Gamma\right)
\eeq where $p\,'$ denotes the commoving particle momentum, with $p\,' \simeq p\,'\,^0$ for 
the energetic particles considered here, $\tau \simeq \lambda/c$ is the mean scattering 
time,- and $\Gamma$ is the shear coefficient. In the strong scattering limit (i.e., $\omega_g\,'
\,\tau \ll 1$, with $\omega_g\,'$ the relativistic gyrofrequency measured in the commoving 
frame), we can use \citep[see][eq.~3.34]{web89}
\beq\label{viscous}
       \Gamma =\frac{c^2}{30}\,\sigma_{\alpha\,\beta}\,
       \sigma^{\alpha\,\beta}\,,
\eeq where $\sigma_{\,\alpha\,\beta}$, with $\alpha,\beta=0,1,2,3$, is the (covariant)
fluid shear tensor given by
\beq\label{shear}
       \sigma_{\,\alpha\,\beta}=\nablaa u_{\beta}+\nablab u_{\alpha}
                 +\dot{u}_{\alpha} u_{\beta}+\dot{u}_{\beta} u_{\alpha}
                 +\frac{2}{3}\left(g_{\,\alpha\,\beta}+u_{\alpha}\,u_{\beta}
                  \right)\,\nabla_{\delta} u^{\delta}\,.
\eeq Here $g_{\alpha \beta}$ denotes the (covariant) metric tensor, e.g., for 
cylindrical coordinates $x^{\alpha} = (c\,t,r,\phi,z)$ we have $(g_{\alpha \beta})
=\rm{diag}(-1,1,r^2,1)$ \citep[e.g.,][]{rie02}, and $\nabla_{\alpha}$ denotes the 
covariant derivative. Note that for cylindrical coordinates, the only non-vanishing 
connection coefficients (Christoffel symbols of 2nd order) are 
\beq
     \Gamma^1_{22}=-r\,, \quad \Gamma^2_{21}=\Gamma^2_{12}=\frac{1}{r}\,.
\eeq
For a time-independent, relativistic bulk flow velocity profile of the form
\beq
           u^{\alpha}= \gamma_b\,\left(1,0,0,\frac{v_z(r)}{c}\right)\,,
\eeq with $\gamma_b(r)=1/\sqrt{1-v_z(r)^2/c^2}$ the bulk Lorentz factor and $u_{\alpha}= 
\gamma_b\,(-1,0,0,v_z(r)/c)$ both, the fluid four divergence, i.e., $\nablab u^{\beta}$, 
and the fluid four acceleration $\dot{u}_{\alpha} \equiv u^{\beta}\,\nablab\,u_{\alpha}$ 
vanish. Moreover, it is easy to show that the only non-vanishing components of the shear 
tensor are given by
\beqn\label{shear-components1}
      \sigma_{01}&=&\sigma_{10} = \nabla_1\,u_0 = \left(\frac{\pad u_0}{\pad x^1} -
                           \Gamma^{\mu}_{01}\,u_{\mu}\right)
                                   = - \frac{\pad \gamma_b(r)}{\pad r}\,,\\
      \sigma_{13}&=&\sigma_{31} =\nabla_1\,u_3 = \frac{\pad u_3}{\pad x^1}
                          = \frac{\pad}{\pad r}\left(\gamma_b(r)\,\frac{v_z(r)}{c}\right)\,
\eeqn Noting that $ \sigma_{01}=-\sigma^{01}$ and $\sigma_{13}= \sigma^{13}$ the 
relativistic shear coefficient could then finally be written as
\beq\label{shear-coeff}
        \Gamma = \frac{1}{15}\,\gamma_b(r)^2 \left(\frac{\pad v_z(r)}{\pad r}\right)^2\,
                 \left[1 + \gamma_b(r)^2\,\frac{v_z(r)^2}{c^2}\right]\,,
\eeq which, for nonrelativistic flow speed (i.e., $\gamma_b \rightarrow 1$ and $v_z/c 
\rightarrow 0$) reduces to the nonrelativistic shear flow coefficient \citep[e.g.,][]{ear88}. 
For an energy-dependent scattering timescale of the form $\tau \propto p\,'\,^{\alpha}$ the
characteristic acceleration timescale can be expressed as
\beq
      t_{\rm sh,l}(r)  \simeq  \frac{p\,'}{<\dot{p}\,'>}  =  \frac{15}{(4+\alpha)}\,
            \frac{c}{\lambda}\,\frac{1}{\gamma_b(r)^2\,\left(\frac{\pad v_z}{\pad r}\right)^2
                      \left[1+\gamma_b(r)^2\,\frac{v_z^2}{c^2}\right]}
\eeq where $v_z \equiv v_z(r)$, and where in the presence of a background magnetic field
along the jet axis $\lambda \equiv \lambda(\gamma)$ has to be smaller than the gyroradius 
to allow for isotropic diffusion characterized by $\kappa_{\perp} \simeq \kappa_{||}$, where 
\citep[e.g.,][eq.~3.7]{web89}
\beq \label{diffusion}
            \kappa_{||}  =  \frac{1}{3}\,v'\,\tau_c\, \quad \quad {\rm and} \quad \quad
            \kappa_{\perp}  =  \frac{\kappa_{||}}{(1+\omega_g '^2\,\tau_c^2)}
\eeq denote the parallel and perpendicular diffusion coefficients.

\clearpage

\end{document}